\documentclass[11p]{elsarticle}
\usepackage{amsmath, amsthm, amssymb, amsfonts}
\usepackage{dsfont}
\usepackage{tikz}
\usepackage{float}

\usepackage{lineno,hyperref}
\modulolinenumbers[5]

\usepackage{caption}
\usepackage{subcaption}

\large\normalsize

\usepackage{graphicx}
\usepackage[mathscr]{eucal}
\usepackage{color}
\journal{Dynamic Games And Applications}

\title{Aspiration can promote cooperation in well-mixed populations as in regular graphs}
\author{Dhaker Kroumi\footnote{Author for
correspondence: dhaker.kroumi@kfupm.edu.sa}
\\
Department of Mathematics and Statistics \\
King Fahd University of Petroleum and Minerals, Dhahran 31261, Saudi Arabia\\
 }

\begin{document}
\maketitle

\section*{Abstract}
Classical studies on aspiration-based dynamics suggest that dissatisfied individuals switch their strategies without taking into account the success of others. The imitation-based dynamics allow individuals to imitate successful strategies without taking into account their own-satisfactions. In this article, we propose to study a dynamic based on aspiration, which takes into account imitation of successful strategies for dissatisfied individuals. Individuals compare their success to their aspired levels. This mechanism helps individuals with a minimum of self-satisfaction to maintain their strategies. Dissatisfied individuals will learn from their neighbors by choosing the successful strategies. We derive an exact expression of the fixation probability in well-mixed populations as in graph-structured populations. As a result, we show that selection favor the evolution of cooperation if the difference in aspired level exceeds some crucial value. Increasing the aspired level of cooperation should oppose cooperative behavior while increasing the aspired level of defection should promote cooperative behavior.  We show that the cooperation level decreases as the connectivity increases. The best scenario for the cooperative evolution is a graph with a small connectivity while the worst scenario is a well-mixed population.

\noindent \textbf{Keywords and phrases}:  Fixation probability; Evolutionary game dynamics; Pair approximation; Cooperation;  Imitation; Aspiration

\noindent \textbf{Mathematics Subject Classification (2010)}: Primary 92D25; Secondary 60J70

\section{Introduction}
Evolutionary game theory is the framework where the frequency of a strategy depends on the fitnesses of the different individuals in the population (Maynard Smith and Price \cite{MP1973}, Maynard Smith \cite{M1982}, Hofbauer and Sigmund \cite{HS1988}, Weibull \cite{W1995}, Samuelson \cite{S1997}, Cressman \cite{C2003}, Vincent and Brown \cite{VB2005}, Nowak \cite{N2006}). Individuals interact and gain payoffs, which are seen as biological fitness or reproductive rates.

The standard model, called the replicator equation, was formulated in an infinitely large well-mixed population where any two individuals have the same probability to interact (Taylor and Jonker \cite{TJ1978}, Zeeman \cite{Z1980}, Hofbauer and Sigmund \cite{HS1998,HS2003}). Suppose that there are $n$ strategies $\{S_1,S_2,\ldots,S_n\}$. The game is described by a payoff matrix $A=\{a_{i,j}\}_{i,j=1,\ldots,n}$, where $a_{i,j}$ is the payoff of an $S_i$-player if its partner is $S_j$-player. Let $x_i$ be the frequency of $S_i$-players in the population. The dynamic is 
\begin{equation}
\frac{dx_i}{dt}=x_i(f_i-\overline{f}),
\end{equation}
where $f_i=\sum\limits_{j=1}^{n}x_ja_{i,j}$ and $\overline{f}=\sum\limits_{i=1}^{n}x_if_{i}$ refer to the expected payoff on an $S_i$-player, and the average payoff in the population, respectively.

Real populations are finite and deterministic approaches cannot capture this finiteness. Recently, a stochastic approach is introduced to model this finiteness by a Markov chain with a finite state space. In the absence of mutation, the Markov chain has absorbing states represented by a population of a unique type. A strategy is said to be favored by selection if its fixation probability is greater than what it would be under neutrality (Nowak \textit{et al.} \cite{NSTF2004}, Imhof and Nowak \cite{IN2006}). In the presence of symmetric mutation, this Markov chain is irreducible, and as a result, it has a stationary state. An interest in the abundance of a given strategy in this equilibrium states arises. In this case, a strategy is said to be favored by selection if its average frequency in the stationary state is greater than what it would be under neutrality (absence of selection) (Antal \textit{et al.} \cite{ANT2009}). Both models, without mutation and with mutation, share the same favored strategy if the mutation rate is small enough (Rousset and Billiard \cite{RB2000}, Rousset \cite{R2003}, Fudenberg and Imhof \cite{FI2006}).

Further advances in evolutionary game theory study structured populations. The traditional setting is the island model where individuals are structured into isolated islands (Ladret and Lessard \cite{LL2007}; Lessard \cite{L2011}). Interactions occur only within islands. After reproduction, individuals can migrate or stay in the parent's island. The case of isolation by distance, called stepping stone model, is considered in Rousset and Billiard \cite{RB2000}, and Rousset \cite{R2006}. Islands are numbered $1,2,\ldots,d$ and the migrate rates are $m_{i,i+1}=m_{i,i-1}=m/2$, $m_{i,i}=1-m$, and $0$ otherwise.

In these structured models, individuals share the same neighborhood if they belong to the same group, or they do not have any common neighbor if they belong to two different groups. Evolutionary graph theory is a natural extension to take into account that individuals can share only some of their neighbors (Nowak and May \cite{NM1992}, Hauert and Doebeli \cite{HD2004}, Lieberman \textit{et al.} \cite{LHN2005}, Ohtsuki \textit{et al.} \cite{OHLN2006}, Taylor \textit{et al.} \cite{TDW2007}). It is a powerful framework that includes social networks in the evolutionary process. $N$ individuals occupy $N$ nodes. Each node is linked to $k$ nodes by edges. Each edge indicates who can interact with whom. 

For a graph of degree $k=2$, the evolutionary process is described in many studies (Ohtsuki and Nowak \cite{ON2006}, van Valen and Nowak \cite{VN2012}). The population state is described completely by the frequency of each strategy. A condition, to favor a strategy over another strategy in a finite population, can be derived as in well-mixed populations.

For general degree $k$, the frequencies of the different strategies are not enough to describe the evolutionary process. To simplify the complexity of such a graph, a technique of pair approximation (Matsuda \textit{et al.} \cite{MOSS1992}, Nakamura \textit{et al.} \cite{NMI1997}, Keeling \cite{K1999}, Van Baalen \cite{B2000}) is introduced to study the evolutionary process in  regular graphs (Ohtsuki \textit{et al.} \cite{ON20062,OHLN2006, OPN2007}). Assume that each individual can choose a strategy among $\{A, B\}$. Pair approximation is a framework to study the stochastic dynamics not only by considering the global frequencies $p_X$ for $X\in\{A, B\}$, but also by considering $q_{X|Y}$ for $X\in\{A, B\}$, the probability that a neighbor of a $Y$-player, is of type $X$. This method assumes that a two-step adjacent player does not affect the focal site directly, that is 
$q_{X|YZ}=q_{X|Y}$. This technique is limited to a very large population such that $k<<N$.

Besides, update rules, in which individuals correct their strategies following a selection mechanism, are of greater importance for their confirmed impact on the evolutionary process. For this reason, one of the most open questions is how do individuals update their  strategies based on their knowledge of themselves and others.

Many update rules have been proposed. The most used are based in two representative models: imitation-based rule (Szab\'o and T\"{o}ke \cite{ST1998}, Ohtsuki \textit{et al.} \cite{OHLN2006}, Traulsen \textit{et al.} \cite{TPN2007}) and aspiration-based rule (Chen and Wang \cite{CW2008}, Matjaz and Zhen \cite{MZ2010}, Du \textit{et al.} \cite{DWW2014,DWAW2014}, Liu \textit{et al.} \cite{LHKP2016}). Under imitation-based rule, individuals update their strategies based on their knowledge of others. An individual compares its payoff with its neighbors' payoffs. If its payoff is lower, it would imitate its neighbors with a higher probability. Under aspiration-based rule, individuals update their strategies based on their knowledge about themselves. An individual compares its payoff to an aspired level, which represents its tolerance with its current strategy. If its payoff is lower, it would switch its strategy with a higher probability.

All these studies suggest that individuals correct their strategies according to only one of the following conceptions: their knowledge of others or their knowledge of themselves. However, real biological species can change their strategies using both information due to the influence of environmental factors and the complexity of their knowledge.
In search of food, foragers of ants use chemical pheromone trails to guide other ants to the food sources. Experienced ants choose to follow the route to their previous trips (Matjaz and Zhen \cite{MZ2010}, Gr\"uter \textit{et al.} \cite{GCR2011}).  Non-experienced ants will imitate their neighbors. This suggests that if the strategy gives the player a certain level of self-satisfaction, then it will be maintained. If individuals did not reach their aspired levels, they will imitate their neighbors. The same conclusion was inferred in experiments on fish stickleback (van Bergen \textit{et al.} \cite{VCL2004}). 

In light of this conclusion, this paper studies the effect of a mixed update rule on the evolution of cooperation in different topologies.  
The update rule is composed of two rounds. In the first round, individuals compares their payoffs to a personal tolerance index. Satisfied individuals keep their current strategy with higher probability. Dissatisfied individuals will observe the success of their neighbors to make a decision. More precisely, a selected individual $I$ compares its payoff $\Pi_I$ to its aspired level $\alpha_I$. It will maintain its current strategy with probability proportional to its satisfaction measured by $\Pi_{I}-\alpha_I$. Otherwise, it will imitate a neighbor's strategy. It will select a neighbor, say $J$, with probability proportional to the fitness of individual $J$. We analyze this model in both well-mixed and graph-structured populations.

This model is equivalent to the following death-birth update. At each time step, a randomly chosen individual, say $I$, survives with probability proportional to $\Pi_{I}-\alpha_I$. Otherwise, it dies. In this case, a competition between its neighbors arises. A neighbor is chosen proportional to its fitness to produce a copy, which will occupy the vacant position. It is similar to the death-birth update rule (Ohtsuki \textit{et al.} \cite{OHLN2006,ON20062}), where the death event occurs with probability $1$.

For a finite well-mixed population (appendix $C$) and a finite graph-structured population of degree $k=2$ (appendix $B$), an exact calculation technique will be used to measure the success of cooperation. We use a property of a discrete Markov chain with two absorbing states to derive the fixation probabilities of cooperation and defection. However, for a graph of degree $k\geq3$, it is not possible to study analytically the evolutionary process in a finite population. However, for a large population, we use a pair approximation technique and then a diffusion approximation to derive the fixation probabilities of cooperation and defection (appendix $A$). 

The remainder of this paper is divided in $4$ sections. In Section $2$, we describe our model. In Section $3$, we test the success of cooperation and defection in well-mixed and graph-structured populations. We apply our results to the simplified additive Prisoner's Dilemma in Section $4$. We finish this article by a discussion in Section $5$.


\section{Model}
Consider a finite population composed of $N$ individuals distributed over $N$ nodes of a graph.  Each node is related by edges to other $k$ nodes. $k$, called the graph degree or the connectivity, is the same for all individuals (see figure $1$). Each edge indicates who interacts with whom. Any two individuals who are related by an edge are called neighbors. Suppose that the graph is connected in the sense that any two nodes are linked by a finite number of edges. Each individual can adopt a strategy among $\{C,D\}$: $C$ for cooperation and $D$ for defection. 

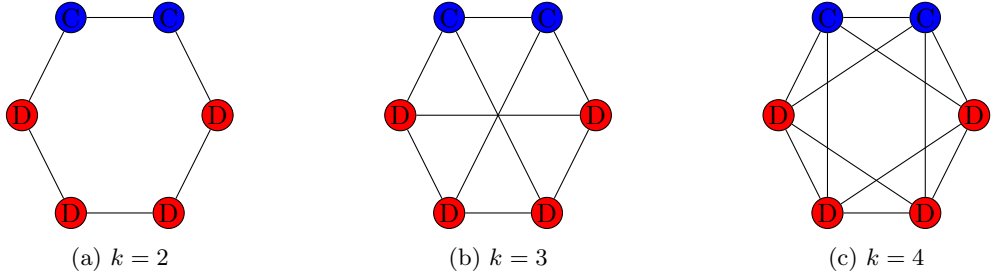
\begin{figure}[h]
       \begin{subfigure}[b]{0.33\linewidth}
		\centering
			\begin{tikzpicture}[auto, scale=1.3]
\tikzstyle{vertex}=[draw, circle, inner sep=0.2mm]
\node (v1) at (0,0) [vertex,fill=blue] {C};
\node (v2) at (1,0) [vertex,fill=blue] {C};
\node (v3) at  (1.5,-1) [vertex,fill=red] {D};
\node (v4) at (1,-2) [vertex,fill=red] {D};
\node (v5) at (0,-2) [vertex,fill=red] {D};
\node (v6) at (-.5,-1) [vertex,fill=red] {D};
\draw (v1) to (v2);
\draw (v2) to (v3);
\draw (v3) to (v4);
\draw (v4) to (v5);
\draw (v5) to (v6);
\draw (v6) to (v1);
\end{tikzpicture}		 
 \caption{$k=2$}
	\end{subfigure}
	\begin{subfigure}[b]{0.33\linewidth}
		\centering
			\begin{tikzpicture}[auto, scale=1.3]
\tikzstyle{vertex}=[draw, circle, inner sep=0.2mm]
\node (v1) at (0,0) [vertex,fill=blue] {C};
\node (v2) at (1,0) [vertex,fill=blue] {C};
\node (v3) at  (1.5,-1) [vertex,fill=red] {D};
\node (v4) at (1,-2) [vertex,fill=red] {D};
\node (v5) at (0,-2) [vertex,fill=red] {D};
\node (v6) at (-.5,-1) [vertex,fill=red] {D};
\draw (v1) to (v2);
\draw (v1) to (v4);
\draw (v2) to (v3);
\draw (v2) to (v5);
\draw (v3) to (v4);
\draw (v3) to (v6);
\draw (v4) to (v5);
\draw (v5) to (v6);
\draw (v6) to (v1);
\end{tikzpicture}		 
 \caption{$k=3$}
	\end{subfigure}
	\begin{subfigure}[b]{0.33\linewidth}
		\centering
	\begin{tikzpicture}[auto, scale=1.3]
\tikzstyle{vertex}=[draw, circle, inner sep=0.2mm]
\node (v1) at (0,0) [vertex,fill=blue] {C};
\node (v2) at (1,0) [vertex,fill=blue] {C};
\node (v3) at  (1.5,-1) [vertex,fill=red] {D};
\node (v4) at (1,-2) [vertex,fill=red] {D};
\node (v5) at (0,-2) [vertex,fill=red] {D};
\node (v6) at (-.5,-1) [vertex,fill=red] {D};
\draw (v1) to (v2); 
\draw (v1) to (v3);
\draw (v2) to (v3);
\draw (v2) to (v4);
\draw (v3) to (v4);
\draw (v3) to (v5);
\draw (v4) to (v5);
\draw (v4) to (v6);
\draw (v5) to (v6);
\draw (v5) to (v1);
\draw (v6) to (v1);
\draw (v6) to (v2);
\end{tikzpicture}
 \caption{ $k=4$}
	\end{subfigure}\hfill
\caption{Each individual is related exactly to $k$ neighbors. $k$ is the same for all individuals, which is called the graph degree. Red nodes are occupied by defectors while blue nodes are occupied by cooperators.}
	\label{interaction+replacement}
\end{figure}

At each time step, each individual interacts with its neighbors through the game matrix
\begin{equation}\label{M1}
\bordermatrix {
& C & D \cr
C& R & S \cr
D & T & P \cr}.
\end{equation} 
Two cooperators receive a reward, $R$, whereas two defectors receive a punishment, $P$.  If they interact, a cooperator receives a sucker, $S$, while a defector receives a temptation, $T$. After interactions with its neighbors, any individual accumulates a payoff denoted by $\Pi$. Then, a randomly chosen individual $I$ will compare its payoff $\Pi_I$ to its satisfaction index $\alpha_I$, which represents its tolerance with its current strategy. Here, we study the simplest case where the satisfaction level is a random variable that depends on the strategy of the individual and does not depend on time. In addition, we assume that $\alpha_I$ is bounded, such that 
\begin{equation}
\mathbb{E}[\alpha_I]=\tilde{\alpha}_I.
\end{equation}
This assumption makes sense because individuals in the population are generally heterogeneous. Therefore, there is an heterogeneity of the aspired level.

 Individual $I$ will keep its current strategy with probability 
\begin{equation}\label{prob-keep}
g\Big(\delta(\Pi_I-\alpha_I)\Big),
\end{equation}
where $\delta$ is a non-negative constant called the selection intensity. It will update its current strategy with the complementary probability 
\begin{equation}\label{prob-imitate}
1-g\Big(\delta(\Pi_I-\alpha_I)\Big).
\end{equation}
In this case, it adopts the strategy of one of its neighbors, say $J$, chosen with probability proportional to its fitness $f_J=1+\delta\Pi_J$. More precisely, let individuals $J_1,J_2,\ldots,J_k$ the neighbors of individual $I$. Individual $I$ will adopt the strategy of one of a neighbor $J_i$ with probability $f_{J_i}/\sum_{l=1}^{k}f_{J_l}$, for $i=1,\ldots,k$. This mechanism helps individuals to learn from their neighbors by selecting the most successful strategies. 

Here, $g$ is a function such that
\begin{itemize}
\item $g(0)=1/2$: for $\delta=0$, updating and maintaining occur with the same probability, that is $1/2$.
\item $g'(0)>0$: for $\delta>0$ very small, maintaining occurs proportional to the satisfaction of individual $I$, which is measured by $\Pi_I-\alpha_I$, since we have
$$g\Big(\delta(\Pi_I-\alpha_I)\Big)\approx\frac{1}{2}+\delta\cdot g'(0)(\Pi_I-\alpha_I).$$
\end{itemize}
If $\Pi_I>\alpha_I$, individuals maintain their strategies with a probability higher than $1/2$. If $\Pi_I<\alpha_I$, individuals maintain their strategies with a probability lower than $1/2$. If $\Pi_I=\alpha_I$, individuals maintain their strategy with probability $1/2$. The most used function is the Fermi rule
\begin{equation}\label{Fermi}
g(x)=\frac{1}{1+e^{-x}}.
\end{equation}
In the remainder, we use the Fermi rule (\ref{Fermi}), where $g'(0)=1/4>0$. 

Neutral model is the case $\delta=0$. The case of weak selection corresponds to $\delta>0$ very small. This case is presented in most studies in genetic populations, molecular evolution and cultural evolution (Kimura \cite{K1983}), Traulsen \textit{et al.} \cite{TPN2007,TSSKM2010}, Wu \textit{et al.} \cite{WAWT2010}). In this case, the effect of payoff differences on the evolutionary process is small. Weak selection is a reasonable assumption for two reasons:
\begin{itemize}
\item It is a standard case to derive many analytic results which are not possible for any selection intensity, but these results stay a good approximation for other selection intensities
\item  In real biological populations, the fitness of an individual depends on many competitions (games), and then each game makes a small contribution, and here we interested only by a game.
\end{itemize} 
In the remainder, we are interested in the effect of weak selection on the evolutionary process. 


\section{Fixation probabilities}

Suppose that a cooperation introduced as a single in an all defecting population. As a result, there are two possibilities for the evolutionary dynamics. The first scenario is that this individual produces a lineage, which will eventually invade the entire population (extinction of defection). The second scenario is that this individual might die before reproducing or generate a lineage that disappears after sometime (extinction of cooperation). The probability of the first scenario, denoted by $\rho_C(\delta)$, is called the fixation probability of cooperation. Similarly, the fixation probability of defection is the probability that a single defector introduced in an all cooperating population produces a lineage, which will take over the population. This probability is denoted by $\rho_C(\delta)$.

\subsection{First test: $\rho_C(\delta)>\rho_C(0)$}

A first criterion, for weak selection to favor the emergence and stabilization of cooperation, is the comparison of the fixation probability under weak selection, $\rho_C(\delta)$, to what it would be under neutrality, $\rho_C(0)=N^{-1}$ (Rousset and Billiard \cite{RB2000}, Nowak \textit{et al.} \cite{NSTF2004}, Taylor \textit{et al.} \cite{TFSN2004}). We say that selection favors the evolution of cooperation if $\rho_C(\delta)>N^{-1}$. Otherwise, that is $\rho_C(\delta)<N^{-1}$, we say that selection disfavors the evolution of cooperation.

\subsubsection{Case 1: $k\geq 3$}
For a finite graph-structured population, the complexity of the graph makes it impossible to study analytically the fixation probability. To simplify such complexity, we use a pair approximation technique to describe the evolutionary process not only by the frequency of cooperators but with the frequency of neighbors of type $CC$. This technique is valid only for a large population such that $3\leq k<<N$. See appendix $A$ for more details.

Using Eq (\ref{sec03-eq8}) in appendix $A$, we have
\begin{equation}\label{sec04-eq1}
\begin{split}
&\rho_C(\delta)=\frac{1}{N}+\delta\cdot\frac{N-1}{12Nk^2}\Big[\Gamma_1+3\Gamma_0+3k^2\Delta\alpha\Big]+\mathcal{O}(\delta^2),
\end{split}
\end{equation}
where
 \begin{equation}
\begin{split}
&\Delta\alpha=\tilde{\alpha}_D-\tilde{\alpha}_C,\\
    &  \Gamma_0=(3k+2)R+(3k^2-3k-2)S-(k+2)T-(3k+2)(k-1)P, \\
      &  \Gamma_1=(3k+2)(k-2)(R-S-T+P).
     \end{split} 
\end{equation}
According to Eq. (\ref{sec04-eq1}), weak selection favors the evolution of cooperation if $\Gamma_0+3\Gamma_1+3k^2\Delta\alpha>0$, which is equivalent to
\begin{equation}\label{sec04-eq2}
\frac{3k^2+5k+2}{3k^2}R+\frac{6k^2-5k-2}{3k^2}S- \frac{3k^2-k+2}{3k^2}T-\frac{6k^2+k-2}{3k^2}P+\Delta\alpha>0.
\end{equation} 
This condition is valid for $k\geq 3$.

\subsubsection{Case 2: $k=2$}
For the circular model $k=2$, we use an exact calculation technique that is valid for any finite population of size $N\geq 3$. See appendix $B$ for details. Using Eq. (\ref{B-eq10}) in appendix $B$, the fixation probability of cooperation is
\begin{equation}\label{sec04-eq3}
\begin{split}
\rho_C(\delta)=&\frac{1}{N}+\delta\cdot\frac{1}{4N^2}\Big[(2N^2-7N+7)R+(N^2+2N-3)S-(N^2-2N+5)T\\
&-(2N^2-3N-5)P+N(N-1)\Delta\alpha\Big]+\mathcal{O}(\delta^2).
\end{split}
\end{equation}
Therefore, weak selection favors the evolution of cooperation if
\begin{equation}
\frac{2N^2-7N+7}{N(N-1)}R+\frac{N^2+2N-3}{N(N-1)}S-\frac{N^2-2N+5}{N(N-1)}T-\frac{2N^2-3N-5}{N(N-1)}P+\Delta\alpha>0.
\end{equation}
For large population, $N\rightarrow\infty$, this condition is reduced to
\begin{equation}
2R+S-T-2P+\Delta\alpha>0,
\end{equation}
which extends Eq. (\ref{sec04-eq2}) for $k=2$.

\subsubsection{Case 3: well-mixed populations}

In a well-mixed population, each individual is connected to all other individuals. By an exact calculation technique, we derived the expression of the probability of fixation of the cooperation for any finite population of size  $N\geq2$. See appendix $C$ for more details. 

Using Eq. (\ref{C-eq5}) in appendix $C$, we have
\begin{equation}\label{sec04-eq4}
\rho_C(\delta)=\frac{1}{N}+\delta\cdot\frac{N-1}{4}\Bigg[\frac{3N-5}{3(N-1)^2}\Big((N-2)R+(2N-1)S-(N+1)T-(2N-4)P\Big)+\Delta\alpha\Bigg]
\end{equation}
As a consequence, weak selection favors the evolution of cooperation if 
\begin{equation}
\frac{3N-5}{3(N-1)^2}\Big((N-2)R+(2N-1)S-(N+1)T-(2N-4)P\Big)+\Delta\alpha>0.
\end{equation}
For a large population, $N\rightarrow\infty$, this condition becomes
\begin{equation}
R+2S-T-2P+\Delta\alpha>0.
\end{equation}
Note that this condition is exactly the limit of condition (\ref{sec04-eq2}) as $k\rightarrow\infty$.

\textbf{Conclusion $1$:}  \textit{For a large structured population in a regular graph of degree $k\geq2$, weak selection favors the evolution of cooperation if 
\begin{equation}\label{sec04-eq5}
\frac{3k^2+5k+2}{3k^2}R+\frac{6k^2-5k-2}{3k^2}S- \frac{3k^2-k+2}{3k^2}T-\frac{6k^2+k-2}{3k^2}P+\Delta\alpha>0.
\end{equation} 
This can be extended for well-mixed populations by taking $k\rightarrow\infty$.}

If inequality (\ref{sec04-eq5}) is reversed, weak selection disfavors the evolution of cooperation, $\rho_C(\delta)<N^{-1}$. This does not mean that weak selection favors the evolution of defection. By symmetry, we have

\textbf{Conclusion $2$:}  \textit{For a large structured population in a regular graph of degree $k\geq2$, weak selection favors the evolution of defection  if 
\begin{equation}\label{sec04-eq6}
\frac{6k^2+k-2}{3k^3}R+\frac{3k^2-k+2}{3k^2}S- \frac{6k^2-5k-2}{3k^2}T-\frac{3k^2+5k+2}{3k^2}P+\Delta\alpha<0.
\end{equation} 
This can be extended for well-mixed populations by taking $k\rightarrow\infty$.}

From Eqs. (\ref{sec04-eq1}), (\ref{sec04-eq3}) and (\ref{sec04-eq4}), it is clear that $\rho_C(\delta)$ increases as $\tilde{\alpha}_D$ increases or $\tilde{\alpha}_C$ decreases. By symmetry $\rho_D(\delta)$ increases as $\tilde{\alpha}_D$ decreases or $\tilde{\alpha}_C$ increases.

\textbf{Conclusion $3$:}  \textit{Increasing the mean of the aspired level of cooperation $\tilde{\alpha}_C$ decreases the fixation probability of cooperation, $\rho_C(\delta)$, and increases the fixation probability of defection, $\rho_D(\delta)$. Increasing the mean of the aspired level of defection $\tilde{\alpha}_D$ increases the fixation probability of cooperation, $\rho_C(\delta)$, and decreases the fixation probability of defection, $\rho_D(\delta)$. This is valid for well-mixed populations as for structured populations in a regular graph of degree $k\geq2$.}

\subsection{Second test: $\rho_C(\delta)>\rho_D(\delta)$}
It is possible that weak selection favors the fixation of cooperation and defection or disfavors the fixation of cooperation and defection. As a result, comparing the fixation probability to what it would be under neutrality does not give a complete view of the success of a strategy. Then, a second criterion is introduced (Nowak \textit{et al.} \cite{NSTF2004}), based on the comparison of the fixation probabilities, to measure the most successful strategy. If $\rho_C(\delta)>\rho_D(\delta)$, then the invasion of a single cooperator in an all defecting population is more likely than the invasion of a single defector in an all cooperating population. In such a case, we say that weak selection favors the evolution of cooperation more than the evolution of defection.

\subsubsection{Case 1: $k\geq 3$}

Using Eq. (\ref{ratio}) of appendix $A$ for $p=N^{-1}$, the ratio of the fixation probabilities is reduced to
\begin{equation}\label{ratio}
\frac{\rho_C(\delta)}{\rho_D(\delta)}=1+\delta\cdot\frac{N-1}{4k^2}\Big[\Gamma_1+2\Gamma_0+2k^2\Delta\alpha \Big]+\mathcal{O}(\delta^2),
\end{equation}
Accordingly, we have
$\rho_C(\delta)>\rho_D(\delta)$ if  $\Gamma_1+2\Gamma_0+2k^2\Delta\alpha>0$, which is equivalent to
\begin{equation}\label{sec04-eq10}
\frac{3k+2}{2k}(R-P)+\frac{3k-2}{2k}(S-T)+\Delta\alpha>0
\end{equation}
This equation predicts the success of cooperators more often than defectors. As mentioned before, the technique used to derive the fixation probabilities is valid for $3\leq k<<N$.

\subsubsection{Case 2: $k=2$}
For $k=2$ and by using Eq. (\ref{B-eq11}) in appendix $B$, we have
\begin{equation}
\frac{\rho_C}{\rho_D}(\delta)=1+\frac{\delta}{2}\Big[(2N-5)(R-P)+N(S-T)+N\Delta\alpha\Big]+\mathcal{O}(\delta^2).
\end{equation}
Accordingly, weak selection favors the evolution of cooperation more than the evolution of defection if 
\begin{equation}
(2N-5)(R-P)+N(S-T)+N\Delta\alpha>0.
\end{equation}
For large population size $N\rightarrow\infty$, this condition is equivalent to 
\begin{equation}
2R+S-T-2P+\Delta\alpha>0,
\end{equation}
which extends condition (\ref{sec04-eq10}) for $k=2$.

\subsubsection{Case 3: Well-mixed populations}
An other extension of condition  (\ref{sec04-eq10}) for well-mixed populations is the following. For a population fully connected, we derive the expression of the ratio $\rho_C/\rho_D$ in Eq. (\ref{C-eq6}) in appendix $B$. We have
\begin{equation}
\frac{\rho_C}{\rho_D}(\delta)=1+\delta\cdot\frac{N-1}{2}\Bigg[\frac{3N-5}{2(N-1)^2}\Big((N-2)R+NS-NT-(N-2)P\Big)+\Delta\alpha\Bigg]+\mathcal{O}(\delta).
\end{equation}
Therefore, weak selection favors the evolution of cooperation more than the evolution of defection if
\begin{equation}
\frac{3N-5}{2(N-1)^2}\Big((N-2)R+NS-NT-(N-2)P\Big)+\Delta\alpha>0.
\end{equation}
For a large population, this condition is reduced to 
\begin{equation}
\frac{3}{2}(R+S-T-P)+\Delta\alpha>0.
\end{equation}
This extends condition (\ref{sec04-eq10}) for $k\rightarrow\infty$.

\textbf{Conclusion $4$:}  \textit{For a large structured population in a regular graph of degree $k\geq2$, weak selection favors the evolution of cooperation more than the evolution of defection if 
\begin{equation}\label{sec04-eq11}
\frac{3k+2}{2k}(R-P)+\frac{3k-2}{2k}(S-T)+\Delta\alpha>0.
\end{equation} 
This is can be extended for well-mixed populations by taking $k\rightarrow\infty$.}

For symmetric aspiration $\tilde{\alpha}_D=\tilde{\alpha}_C$, condition (\ref{sec04-eq11}) for weak selection to favor the evolution of cooperation more than the evolution of defection can be written as 
\begin{equation}\label{sec04-eq12}
\sigma R+S>T+\sigma P,
\end{equation}
where $\sigma=\frac{3k+2}{3k-2}$ is the structure coefficient (Tarnita \textit{et al.} \cite{TOAFN2009}).
Here, $\sigma$ describes the structure and the update rule effects on the evolutionary process. It does not depend on the game matrix. It quantifies the degree for which individuals of the same type are more likely to meet than individuals of different types. If we select two neighbors, we have different types with probability $1/(1+\sigma)$, or the same type with probability $\sigma/(1+\sigma)$. 

The game is equivalent to a well-mixed population without structure, where each individual can interact with any other individual through the effective game matrix (Lessard \cite{L2011}), given by
\begin{equation}
A_{eff}=\begin{bmatrix}
    \sigma R & S \\
    T & \sigma P
  \end{bmatrix}.
\end{equation}
Note that $\sigma$ converges to $1$ as $k\rightarrow\infty$. Therefore, the normal payoff matrix (\ref{M1}) is obtained in the limit where each individual interacts with any other individual. This describes exactly a well-mixed population and the limit of condition (\ref{sec04-eq5}) is
\begin{equation}
R+S>T+P.
\end{equation}
This is exactly the limit of the condition for weak selection to favor the evolution of cooperation more than the evolution of defection for a well-mixed population that follows a Moran procedure (Taylor \textit{et al.} \cite{TFSN2004}). In a well-mixed population, the update rule has no effect on the evolutionary process.


\section{Example: the simplified additive Prisoner's Dilemma}
Consider the simplified additive Prisoner's Dilemma given by the matrix
\begin{equation}\label{additive}
\bordermatrix {
& C & D \cr
C& b-c & -c \cr
D & b & 0 \cr}.
\end{equation}
A cooperator pays a cost $c>0$ to receive a benefit $b>c$ if its partner cooperates. A defector benefits by receiving $b$ if its partner cooperates. This is one of the most important social dilemmas, which can be used to quantify the effectiveness of cooperation via the benefit-to-cost ratio $b/c$. This ratio is an indicator of the performance of cooperation in structured populations as in well-mixed populations.

Using condition (\ref{sec04-eq5}) with the new entries, weak selection favors the evolution of cooperation, $\rho_C(\delta)>N^{-1}$, if
\begin{equation}\label{cooperation}
\Delta\alpha+\frac{2}{k}b>3c.
\end{equation}
Note that this condition is exactly condition (\ref{sec04-eq6}) for weak selection to disfavor the evolution of defection, $\rho_D(\delta)<N^{-1}$, and condition (\ref{sec04-eq11}) to favor the evolution of cooperation more than the evolution of defection $\rho_C(\delta)>\rho_D(\delta)$.
Therefore, we cannot have $\rho_C(\delta)>N^{-1}$ and $\rho_D(\delta)>N^{-1}$ or $\rho_C(\delta)<N^{-1}$ and $\rho_D(\delta)<N^{-1}$. The difference in aspired level, $\Delta\alpha=\tilde{\alpha}_D-\tilde{\alpha}_C$, is a form of compensation to cooperators for their behavior. Weak selection fully favors the evolution of cooperation, that is $\rho_C(\delta)>N^{-1}>\rho_D(\delta)$, if the compensation $\Delta\alpha$ exceeds the difference in payoff $3c-2k^{-1}b$. 

Otherwise, that is 
\begin{equation}
\Delta\alpha+\frac{2}{k}b<3c,
\end{equation}
 weak selection fully favors the evolution of defection, that is $\rho_C(\delta)<N^{-1}<\rho_D(\delta)$. In this case, the difference in aspired level, $\Delta\alpha$, is not enough to compensate cooperators to evolve and take over the population. Selection should oppose cooperative behavior.

With large values of $\Delta\alpha$, a cooperator will be more satisfied than a defector. This allows cooperators to maintain their strategy more frequently than defectors and increases the updating frequency of defectors until they finish by accepting cooperation.

The weight of the benefit $b$ on condition (\ref{cooperation}) depends on the connectivity $k$. For a graph with the smallest connectivity, $k=2$, weak selection fully favors the evolution of cooperation if
\begin{equation}\label{k=2}
\Delta\alpha+b>3c.
\end{equation}
For a graph with the largest connectivity $k\rightarrow\infty$ (well-mixed populations), weak selection fully favors the evolution of cooperation, if
\begin{equation}\label{k=infty}
\Delta\alpha>3c.
\end{equation}
For any other connectivity, the condition is between (\ref{k=2}) and (\ref{k=infty}).
The first condition is the least stringent one and the second condition the most stringent one. This suggests that the best scenario for the cooperative evolution is a graph with a small connectivity. Increasing the connectivity reduces the cooperation level.

Consider the case where each type aspires in average the maximum payoff that can receive it, $\tilde{\alpha}_C=b-c$ and $\tilde{\alpha}_D=b$. Then, the difference in aspired level is $\Delta\alpha=c$. Condition (\ref{cooperation}) for selection to fully favor the evolution cooperation becomes 
\begin{equation}
\frac{b}{c}>k.
\end{equation}
This is typically the condition derived by Ohtsuki \textit{et al.} \cite{OHLN2006} for death-birth updating, where at each time step, an individual is selected to die. Then, a neighbor is selected with probability proportional to its fitness to give birth to a copy, which will take the vacant position.
Note that, for a graph with a large connectivity, weak selection fully favors the evolution of defection whatever the benefit $b$ and the cost $c$.

If both strategies aspire in average the same level, that is $\Delta\alpha=0$, then weak selection fully favors the evolution of cooperation if 
\begin{equation}\label{eq34}
\frac{b}{c}>\Big(\frac{b}{c}\Big)^{*}=\frac{\sigma+1}{\sigma-1}=\frac{3}{2}k.
\end{equation}
Decreasing the connectivity $k$ decreases the crucial ratio $\Big(\frac{b}{c}\Big)^{*}$ that $b/c$ should exceed it for weak selection ti favor the evolution of cooperation. These results reveal that larger is the value of the connectivity $k$, larger must be the value of $\Big(\frac{b}{c}\Big)^{*}$ for selection to favor the evolution of cooperation in any sense. For a very large connectivity $k\rightarrow\infty$, weak selection fully favors the evolution of defection.

\textbf{Conclusion $5$:}  \textit{In the case of simplified additive Prisoner's Dilemma and under weak selection, the condition $\rho_C(\delta)>N^{-1}$ is sufficient for $\rho_C(\delta)>N^{-1}>\rho_D(\delta)$. This condition is 
\begin{equation}
\Delta\alpha+\frac{2}{k}b>3c.
\end{equation} 
Increasing the connectivity reduces the cooperation level.}

\section{Discussion}

Strategy update rule, in which individuals correct their strategies following a selection dynamic, is a microscopic mechanism that can serve to explain the cooperative evolution in different topologies. Two fundamentals update rules are the most used: aspiration-based mechanisms, which are based on the knowledge of individuals about themselves (self-learning), and imitation-based mechanisms, which are based on the knowledge of individuals about their neighborhood. To date, studies on evolutionary dynamics have focused on one of these mechanisms.

In this paper, we have established a mixed update rule, in which individuals test their success with their aspired levels to decide whether or not to imitate their neighbors. Individuals in the population are generally heterogeneous. Consequently, we have considered along with this paper a heterogeneity of aspired level, which is a bounded random variable with a mean depends on the strategy. Satisfied individuals will maintain their strategies with a higher probability. Dissatisfied individuals will imitate one of their neighbors with a higher probability. In this case, a strategy is selected with probability proportional to its fitness.

For a general game, we have derived the fixation probabilities of cooperation and defection. For the particular cases, circular model and well-mixed population, we have the exact values of the fixation probabilities for finite population. For a general graph, we have an approximation of the fixation probabilities for a large population.

We applied these results to test the success of cooperation. We have shown that weak selection favors the evolution of cooperation more than the evolution of defection if
\begin{equation}\label{conclusion1}
\frac{3k+2}{2k}(R-P)+\frac{3k-2}{2k}(S-T)+\Delta\alpha>0,
\end{equation}
where $k$ is the graph degree. This condition can be extended to large well-mixed populations by tending $k\rightarrow\infty$. 
$\frac{3k+2}{2k}(R-P)+\frac{3k-2}{2k}(S-T)$ quantifies the effect of the payoff difference between cooperation and defection. $\Delta\alpha$ quantifies the effect of the difference in aspired level between cooperation and defection.

We have shown that an increase in the mean of the aspired level of defection, or a decrease in the mean of the aspired level of cooperation, makes it easier for $\rho_C>N^{-1}$, $\rho_C>\rho_D$ and $\rho_D<N^{-1}$ to hold. The conclusion is that these conditions tend to promote the evolution of cooperation. This is true in well-mixed populations as in graph-structured populations. On the other hand, a decrease in the mean of the aspired level of defection, or an increase in the mean of the aspired level of cooperation tends to oppose the evolution of cooperation.

For symmetric aspiration $\tilde{\alpha}_C=\tilde{\alpha}_D$, weak selection favors the evolution of cooperation more than the evolution of defection if
\begin{equation}\label{conclusion2}
\sigma R+S>T+\sigma P,
\end{equation}
where $\sigma=\frac{3k+2}{3k-2}$ is the coefficient introduced by Tarnita \textit{et al.} \cite{TOAFN2009}. It describes the effect of the structure and the update rule on the evolutionary process. It does not depend on the game matrix. For a well-mixed population, we have $\sigma=1$ and then condition (\ref{conclusion2}) becomes $R+S>T+P$, which is exactly the risk dominance condition in a coordination game (Harsanyi and Selten \cite{HSe1988}).  A coordination game is the case where $R>T$ and $P>S$.

Of further interest is the effect of the mixed dynamic in the additive simplified Prisoner's Dilemma.
The condition for weak selection to favor the evolution of cooperation, $\rho_C(\delta)>N^{-1}$, is sufficient for weak selection to disfavor the evolution of defection, $\rho_D(\delta)<N^{-1}$, and favor the evolution of cooperation more than the evolution of defection, $\rho_C(\delta)>\rho_D(\delta)$. This is true for well-mixed populations as in graph-structured populations.
This conclusion is in agreement with the conclusion obtained for other update rules as Birth-death, death-birth, imitation and pairwise comparison (Nowak \textit{et al.} \cite{NSTF2004}; Nowak \cite{N2006}).
Under our mixed model, this condition is
\begin{equation}\label{conclusion}
\Delta\alpha+\frac{2}{k}b>3c.
\end{equation} 
 The best scenario for the evolution of cooperation is the circular model, $k=2$, where the benefit $b$ has a major effect on the favored strategy. In this case, the condition becomes $\Delta\alpha+b>3c$. It requires that $\Delta\alpha>2c$ since $b>c$.  Increasing the connectivity $k$ reduces the cooperation level since it reduces the weight of the benefit $b$ on condition (\ref{conclusion}). For a large well-mixed population, which corresponds to a graph with a large connectivity $k\rightarrow\infty$, the benefit $b$ does not come into play in this condition. In this case, the condition is $\Delta\alpha>3c$. It is clear that the condition for a graph in a large connectivity, $k\rightarrow\infty$, is the most stringent one and the condition in the circular model, $k=2$, the least stringent one. 

For the particular case, when each type aspires in average the maximum payoff that can receive it, $\tilde{\alpha}_C=b-c$ and $\tilde{\alpha}_D=b$, weak selection favors the evolution of cooperation in any sense if  
\begin{equation}
\frac{b}{c}>k,
\end{equation}
which is typically the condition derived by Ohtsuki \textit{et al.} \cite{OHLN2006} for death-birth updating. 
If both strategies aspire in average the same level, that is $\Delta\alpha=0$, then weak selection fully favors the evolution of cooperation if 
\begin{equation}\label{eq34}
\frac{b}{c}>\frac{3}{2}k.
\end{equation}
In both cases, it is possible for weak selection to favor the evolution of cooperation if the connectivity $k$ is finite and $b$ is sufficiently large. In such a case, cooperators form clusters that emerge over the graph. However, for large connectivity, weak selection always disfavors the evolution of cooperation.

\section*{Acknowledgments}
This research was funded by Deanship of Scientific Research (DSR) at King Fahd University of Petroleum and Minerals (Grant No. $SR181014$).

\section*{References}
\bibliographystyle{unsrt}

\newpage

\section{Appendix A: General case $k\geq3$}
\subsection{Rate of change under weak selection}
Define $p_X$ and $p_{XY}$ as the frequencies of strategy $X$ and pairs of type $XY$, respectively, for $X,Y\in\{C,D\}$. Denote by $q_{Y|X}=p_{XY}/p_X$ the probability that a given neighbor of an $X$-strategist is a $Y$-strategist. As a result of basic probability properties, these quantities are related by the following relations
\begin{align}
&p_C+p_D=1,\nonumber\\
&p_{XY}=q_{X|Y}p_Y=q_{Y|X}p_X=p_{YX},\\
&q_{A|Y}+q_{B|Y}=1.\nonumber
\end{align}
Using these identities, we can express all these probabilities in terms of $p_C$ and $q_{C|C}$ 
\begin{equation}\label{sec02-eq1}
\begin{split}
p_D&=1-p_C,\\
p_{CC}&=q_{C|C}p_C,\\
q_{D|C}&=1-q_{C|C},\\
p_{CD}&=q_{D|C}p_C=(1-q_{C|C})p_C,\\
q_{C|D}&=\frac{p_{CD}}{p_D}=\frac{(1-q_{C|C})p_C}{1-p_C},\\
q_{D|D}&=1-q_{C|D}=1-\frac{(1-q_{C|C})p_C}{1-p_C},\\
p_{DD}&=q_{D|D}p_D=1-p_C-(1-q_{C|C})p_C.
\end{split}
\end{equation}
Based on the above identities, the evolutionary process is completely described through $p_C$ and $q_{C|C}$. 

The next step is to derive the changes in one time step of $p_C$ and $q_{C|C}$, respectively, to characterize the evolutionary process of our model.


\subsubsection{Payoffs}

Assume that the selected individual, $I$, is an $X$-strategist, and that its neighborhood is formed by $k_C$ cooperators and $k_D=k-k_C$ defectors. Therefore, its expected payoff is 
\begin{equation}\label{sec02-eq2}
\Pi_X = \left\{
    \begin{array}{ll}
       \frac{k_CR+k_DS}{k} & \mbox{if } X=C ,\\
         \frac{k_CT+k_DP}{k} & \mbox{if } X=D. \\
    \end{array}
\right.
\end{equation}

In the second round, individual $I$ will adopt a neighbor's strategy. Then, we must consider the neighborhood's payoffs of $I$. Let individual $J$, a $Y$-player, be a random neighbor of individual $I$, and let $\Pi_{Y|X}$ be the expected payoff of individual $J$. Hence, with reasoning based on the strategies of individuals $I$ and $J$, we have
\begin{equation}\label{sec02-eq3}
\begin{split}
 & \Pi_{C|C}=\frac{[1+(k-1)q_{C|C}]R+(k-1)q_{D|C}S}{k},\\
 &  \Pi_{C|D}=\frac{(k-1)q_{C|C}R+[1+(k-1)q_{D|C}]S}{k},\\
   &  \Pi_{D|C}=\frac{[1+(k-1)q_{C|D}]T+(k-1)q_{D|D}P}{k},   \\
 &   \Pi_{D|D}=\frac{(k-1)q_{C|D}T+[1+(k-1)q_{D|D}]P}{k}. 
    \end{split}
 \end{equation}
 \begin{proof}
Start with the first payoff in Eq. (\ref{sec02-eq3}). Assume that $I$ and $J$ are two cooperators. In addition of $I$, individual $J$ has other $k-1$ neighbors. Each one of them is of type $C$ with probability $q_{C|C}$, or of type $D$ with probability $q_{D|C}=1-q_{C|C}$. In average, the neighborhood of individual $J$ is composed of $1+(k-1)q_{C|C}$ cooperators and $(k-1)q_{D|C}$ defectors. This explains the form of the expected payoff. Similarly, we have the other payoffs in Eq. (\ref{sec02-eq3}).
\end{proof}


\subsubsection{Change  in $p_C$}
The frequency of $C$, $p_C$, increases if a defector becomes a cooperator. A defector is selected to update its strategy with probability $p_D$. Its neighborhood is formed by $k_C$ cooperators and $k_D=k-k_C$ defectors with probability $\binom{k}{k_C}q_{C|D}^{k_C}q_{D|D}^{k_D}$, for $k_C=0,1,\ldots,k$. It will choose to update its strategy with probability 
\begin{equation}
\mathbb{E}\Big[\frac{1}{1+e^{\delta(\Pi_D-\alpha_D)}}\Big]=\mathbb{E}\Big[\frac{1}{2}+\delta\cdot\frac{\alpha_D-\Pi_D}{4}+\mathcal{O}(\delta^2)\Big]=\frac{1}{2}+\delta\cdot\frac{\tilde{\alpha}_D-\Pi_D}{4}.
\end{equation}
Here $\mathcal{O}(\delta^n)$ means that the error is of order of $\delta^n$ for $n\in\mathbb{N}$. Finally, it becomes a cooperator with probability 
\begin{equation}
\frac{k_C(1+\delta\Pi_{C|D})}{k_C(1+\delta\Pi_{C|D})+k_D(1+\delta\Pi_{D|D})}=\frac{k_C}{k}+\delta\frac{k_Ck_D}{k^2}(\Pi_{C|D}-\Pi_{D|D})+\mathcal{O}(\delta^2).
\end{equation}
In this case, the change is $\Delta p_C=\frac{1}{N}$. Summarize this event in the following probability
\begin{equation}
\begin{split}
\mathbb{P}\Big(\Delta p_C=\frac{1}{N}\Big)=&\underbrace{p_D}_\text{select a defector} \sum_{k_C=0}^{k}\underbrace{\binom{k}{k_C}q_{C|D}^{k_C}q_{D|D}^{k_D}}_\text{ its neighborhood}\underbrace{\mathbb{E}\Big[\frac{1}{1+e^{\delta(\Pi_D-\alpha_D)}}\Big]}_\text{it updates its strategy}\underbrace{\frac{k_C(1+\delta\Pi_{C|D})}{k_C(1+\delta\Pi_{C|D})+k_D(1+\delta\Pi_{D|D})}}_\text{it becomes a cooperator} \\
=&\frac{p_D}{2k}\sum_{k_C=0}^{k}k_C\binom{k}{k_C}q_{C|D}^{k_C}q_{D|D}^{k_D}+\frac{\delta p_D}{2}\sum_{k_C=0}^{k}\frac{k_C}{k}\binom{k}{k_C}q_{C|D}^{k_C}q_{D|D}^{k_D}\\
&\;\;\;\;\;\;\;\;\;\;\;\;\;\;\;\;\;\;\;\;\;\;\;\;\;\;\;\;\;\;\times\Big[ \frac{k_D}{k}(\Pi_{C|D}-\Pi_{D|D})+\frac{\tilde{\alpha}_D-\Pi_{D}}{2}\Big]+\mathcal{O}(\delta^2).
\end{split}
\end{equation}
Using Eq. (\ref{sec02-eq3}) and  the first two moments of the binomial distribution,
\begin{equation}\label{sec02-eq4}
\begin{split}
&\sum_{k_C=0}^{k}k_C\times\binom{k}{k_C}q_{C|D}^{k_C}q_{D|D}^{k_D}=\sum_{k_C=0}^{k}k_C\times\binom{k}{k_C}q_{C|D}^{k_C}(1-q_{C|D})^{k_D}=kq_{C|D},\\
&\sum_{k_C=0}^{k}k_C^2\times\binom{k}{k_C}q_{C|D}^{k_C}q_{D|D}^{k_D}=\sum_{k_C=0}^{k}k_C^2\times\binom{k}{k_C}q_{C|D}^{k_C}(1-q_{C|D})^{k_D}=kq_{C|D}+k(k-1)q_{C|D}^2,
\end{split}
\end{equation}
yield 
\begin{equation}\label{sec02-eq5}
\begin{split}
\mathbb{P}\Big(\Delta p_C=\frac{1}{N}\Big)=&\frac{p_Dq_{C|D}}{2}+\frac{\delta p_D}{4k}\Big[2(k-1)q_{C|D}q_{D|D}(\Pi_{C|D}-\Pi_{D|D})+kq_{C|D}\tilde{\alpha}_D\\
&-(1+(k-1)q_{C|D})q_{C|D}T-(k-1)q_{C|D}q_{D|D}P\Big]+\mathcal{O}(\delta^2)\\
=&\frac{p_{CD}}{2}+\frac{\delta p_{CD}}{4k}\Big[I_{R}^+R+I_{S}^+S-I_{T}^+T-I_{P}^+P+k\tilde{\alpha}_D\Big]+\mathcal{O}(\delta^2),
\end{split}
\end{equation}
where
\begin{equation}\label{sec02-eq6}
\begin{split}
&I_R^+=\frac{2(k-1)^2}{k}q_{C|C}q_{D|D},\\
&I_S^+=\frac{2(k-1)}{k}q_{D|D}\Big(1+(k-1)q_{D|C}\Big),\\
&I_T^+=1+(k-1)q_{C|D}+\frac{2(k-1)^2}{k}q_{D|D}q_{C|D},\\
&I_P^+=\frac{k-1}{k}q_{D|D}\Big(k+2+2(k-1)q_{D|D}\Big).
\end{split}
\end{equation}

The frequency of $C$, $p_C$, decreases if a cooperator becomes a defector. In this case, the change is $\Delta p_C=-\frac{1}{N}$. This happens with probability 
\begin{align}\label{sec02-eq7}
\mathbb{P}\Big(\Delta p_C=-\frac{1}{N}\Big)&=\underbrace{p_C}_\text{select a cooperator } \sum_{k_C=0}^{k}\underbrace{\binom{k}{k_C}q_{C|C}^{k_C}q_{D|C}^{k_D}}_\text{ its neighborhood}\underbrace{\mathbb{E}\Big[\frac{1}{1+e^{\delta(\Pi_C-\alpha_C)}}\Big]}_\text{it updates its strategy}\underbrace{ \frac{k_D(1+\delta\Pi_{D|C})}{k_C(1+\delta\Pi_{C|C})+k_D(1+\delta\Pi_{D|C})}}_\text{it becomes a defector} \nonumber\\
&=\frac{p_C}{2k}\sum_{k_C=0}^{k}k_D\binom{k}{k_C}q_{C|C}^{k_C}q_{D|C}^{k_D}+\frac{\delta p_C}{2k}\sum_{k_C=0}^{k}k_D\binom{k}{k_C}q_{C|C}^{k_C}q_{D|C}^{k_D}\nonumber\\
&\;\;\;\;\;\;\;\;\;\;\;\;\;\;\;\;\;\;\;\;\;\;\;\;\;\;\;\;\;\;\times\Big( \frac{k_C}{k}(\Pi_{D|C}-\Pi_{C|C})+\frac{\tilde{\alpha}_C-\Pi_{C}}{2}\Big)+\mathcal{O}(\delta^2)\nonumber\\
&=\frac{p_{CD}}{2}+\frac{\delta p_{CD}}{4k}\Big(-I_{R}^-R-I_{S}^-S+I_{T}^-T+I_{P}^-P+k\tilde{\alpha}_C\Big)+\mathcal{O}(\delta^2),
\end{align}
where
\begin{equation}\label{sec02-eq8}
\begin{split}
&I_R^-=\frac{k-1}{k}q_{C|C}\Big(k+2+2(k-1)q_{C|C}\Big),\\
&I_S^-=1+(k-1)q_{D|C}+\frac{2(k-1)^2}{k}q_{C|C}q_{D|C},\\
&I_T^-=\frac{2(k-1)}{k}q_{C|C}\Big(1+(k-1)q_{C|D}\Big),\\
&I_P^-=\frac{2(k-1)^2}{k}q_{C|C}q_{D|D}.
\end{split}
\end{equation}

Denote by $\dot{p}_A$ the rate of change of $p_C$ in one time step. Using Eqs (\ref{sec02-eq5}) and (\ref{sec02-eq7}), we obtain
\begin{align}\label{sec02-eq9}
\dot{p}_C=&\frac{1}{N}\mathbb{P}\Big(\Delta p_C=\frac{1}{N}\Big)-\frac{1}{N}\mathbb{P}\Big(\Delta p_C=-\frac{1}{N}\Big)\nonumber\\
=&\frac{\delta p_{CD}}{4Nk}\Big[I_RR+I_SS-I_TT-I_PP+k\Delta\alpha\Big]+\mathcal{O}(\delta^2),
\end{align}
where
\begin{equation}\label{sec02-eq10}
\begin{split}
I_R=I^+_R+I^-_R=&\Big[k+2+2(k-1)(q_{C|C}+q_{D|D})\Big]\frac{k-1}{k}q_{C|C},\\
I_S=I^+_S+I^-_S=&1+(k-1)q_{D|C}+\frac{2(k-1)^2}{k}(q_{C|C}+q_{D|D})q_{D|C}+\frac{2(k-1)q_{D|D}}{k},\\
I_T=I^+_T+I^-_T=&1+(k-1)q_{C|D}+\frac{2(k-1)^2}{k}(q_{C|C}+q_{D|D})q_{C|D}+\frac{2(k-1)q_{C|C}}{k},\\
I_P=I^+_P+I^-_P=&\Big[k+2+2(k-1)(q_{C|C}+q_{D|D})\Big]\frac{k-1}{k}q_{D|D},
\end{split}
\end{equation}
and $\Delta\alpha=\tilde{\alpha}_D-\tilde{\alpha}_C$. $\Delta\alpha$ is the difference in aspired level between a defector and a cooperator.


\subsubsection{Change in $q_{C|C}$}
Since $q_{C|C}=p_{CC}/p_C$, we must start by the rate of change of $p_{CC}$, the frequency of $CC$-pairs. Note that the total number of all pairs is $kN/2$ as each individual has $k$ neighbors. $p_{CC}$ changes if a defector becomes a cooperator or a cooperator becomes a defector. Let $I$ be the selected individual and assume that its neighborhood is formed by $k_C$ cooperators and $k_D=k-k_C$ defectors.

The number of pairs $CC$ will increase by $k_C$ if a defector becomes a cooperator. This occurs with probability
\begin{align}\label{sec02-eq11}
\mathbb{P}\Big(\Delta p_{CC}=\frac{2k_C}{kN}\Big)=&p_D\binom{k}{k_C}q_{C|D}^{k_C}q_{D|D}^{k_D}\times\mathbb{E}\Big[\frac{1}{1+e^{\delta(\Pi_D-\alpha_D)}}\Big]\times \frac{k_C(1+\delta\Pi_{C|D})}{k_C(1+\delta\Pi_{C|D})+k_D(1+\delta\Pi_{D|D})}\nonumber\\
=&\frac{k_Cp_{D}}{2k}\binom{k}{k_C}q_{C|D}^{k_C}q_{D|D}^{k_D}+\mathcal{O}(\delta),
\end{align}
The number of pairs $CC$ will decrease by $k_C$ if a cooperator becomes a defector. This occurs with probability
\begin{align}\label{sec02-eq12}
\mathbb{P}\Big(\Delta p_{CC}=-\frac{2k_C}{kN}\Big)=&p_C\binom{k}{k_C}q_{C|C}^{k_C}q_{D|C}^{k_D}\times\mathbb{E}\Big[\frac{1}{1+e^{\delta(\Pi_C-\alpha_C)}}\Big]\times \frac{k_D(1+\delta\Pi_{D|C})}{k_C(1+\delta\Pi_{C|C})+k_D(1+\delta\Pi_{D|C})}\nonumber\\
=&\frac{p_{C}k_D}{2k}\binom{k}{k_C}q_{D|C}^{k_C}q_{C|C}^{k_D}+\mathcal{O}(\delta).
\end{align}

Let $\dot{p}_{CC}$ be the rate of change of $p_{CC}$ in one time step. Combining Eqs (\ref{sec02-eq4}) and  (\ref{sec02-eq12}) yield
\begin{align}\label{sec02-eq13}
\dot{p}_{CC}=&\sum_{k_C=0}^{k}\frac{2k_C}{kN}\Bigg[\mathbb{P}\Big(\Delta p_{CC}=\frac{2k_C}{kN}\Big)-\mathbb{P}\Big(\Delta p_{CC}=-\frac{2k_C}{kN}\Big)\Bigg]\nonumber\\
=&\sum_{k_C=1}^{k}\frac{2k_C}{kN}\Bigg[\frac{k_Cp_{D}}{2k}\binom{k}{k_C}q_{C|D}^{k_C}q_{D|D}^{k_D}-\frac{k_Dp_{C}}{2k}\binom{k}{k_C}q_{C|C}^{k_C}q_{D|C}^{k_D}\Bigg]+\mathcal{O}(\delta)\nonumber\\
=&\frac{p_{D}}{k^2N}\sum_{k_C=0}^{k}k_C^2\binom{k}{k_C}q_{C|D}^{k_C}q_{D|D}^{k_D}-\frac{p_{C}}{k^2N}\sum_{k_C=0}^{k}k_Ck_D\binom{k}{k_C}q_{C|C}^{k_C}q_{D|C}^{k_D}+\mathcal{O}(\delta)\nonumber\\
=&\frac{p_{D}}{k^2N}\Big(kq_{C|D}-k(k-1)q_{C|D}^2\Big)-\frac{p_{C}}{k^2N}\Big(k^2q_{C|C}-kq_{C|C}-k(k-1)q_{C|C}^2\Big)+\mathcal{O}(\delta)\nonumber\\
=&\frac{p_{CD}}{kN}\Big(1+(k-1)(q_{C|D}-q_{C|C})\Big)+\mathcal{O}(\delta).
\end{align}
As a result, the rate of change of $q_{C|C}$ in one time step is
\begin{align}\label{sec02-eq14}
\dot{q}_{C|C}=&\frac{\dot{p}_{CC}}{p_C}=\frac{p_{CD}}{kNp_C}\Big(1+(k-1)(q_{C|D}-q_{C|C})\Big)+\mathcal{O}(\delta)\nonumber\\
&=\frac{1-q_{C|C}}{kN}\Big\{1+(k-1)\frac{p_C-q_{C|C}}{1-p_C}\Big\}+\mathcal{O}(\delta).
\end{align}
In the last step, we have used Eq (\ref{sec02-eq1}).



\subsection{The quasi-steady state}
Under weak selection, global frequency $p_C$ changes at a rate of order $\delta$ (see Eq. (\ref{sec02-eq9})), which is very small, while the local frequency $q_{C|C}$ changes at a rate of order $1$ (see Eq. (\ref{sec02-eq14})). As a consequence, the local density $q_{C|C}$ equilibrates much more quickly than the global density $p_C$ (Ohtsuki and Nowak \cite{ON2006}). Therefore, the dynamical system rapidly converges onto a quasi-steady state, defined by $\dot{q}_{C|C}=0$, or more explicitly, 
\begin{equation}\label{sec03-eq1}
q_{C|C}=\frac{1}{k-1}+\frac{k-2}{k-1}p_C.
\end{equation}
It is the key relationship, which is obtained in many studies of structured populations in regular graphs (Ohtsuki \textit{et al.} \cite{OHLN2006,ON2006,OPN2007}). 

Instead of studying a diffusion process in terms of two variables, $p_C$ and $q_{C|C}$, by this relation we describe the system by one-dimensional diffusion process in terms of $p_C$ only. In fact, by using Eq (\ref{sec03-eq1}), we express the different probabilities of Eq (\ref{sec02-eq1}) in terms of $p_C$ as 
\begin{equation}\label{A-eq1}
\begin{split}
p_D=&1-p_C,\\
q_{D|C}=&1-q_{C|C}=\frac{k-2}{k-1}(1-p_C),\\
q_{C|D}=&\frac{q_{D|C}p_C}{1-p_C}=\frac{k-2}{k-1}p_C,\\
q_{D|D}=&1-q_{C|D}=1-\frac{k-2}{k-1}p_C,\\
p_{CD}=&q_{D|C}p_C=\frac{k-2}{k-1}(1-p_C)p_C.
\end{split}
\end{equation}

Inserting these equations in Eq (\ref{sec02-eq10}), we have 
\begin{equation}\label{A-eq3}
\begin{split}
I_R=&\frac{3k+2+(3k+2)(k-2)p_C}{k},\\
I_S=&\frac{3k^2-3k-2-(k-2)(3k+2)p_C}{k},\\
I_T=&\frac{(k+2)+(3k+2)(k-2)p_C}{k},\\
I_P=&\frac{(3k+2)(k-1)-(3k+2)(k-2)p_C}{k}.
\end{split}
\end{equation}

With a short interval $\Delta t$ and by using Eqs (\ref{sec02-eq9}) and (\ref{A-eq3}), we have
\begin{equation}\label{sec03-eq2}
\mathbb{E}\Big[\Delta p_C\Big|p_C(0)=p\Big]=\dot{p}_{C}\Delta t\Bigg(\equiv \mu(p)\Delta t\Bigg).
\end{equation}
Here $\mu$ is the first order given by
\begin{equation}\label{sec03-eq2}
\mu(p)=\frac{\delta(k-2)}{4Nk^2(k-1)}p(1-p)\Big(\Gamma_0+k^2\Delta\alpha+\Gamma_1p\Big),
\end{equation}
where 
\begin{equation}\label{sec03-eq3}
\begin{split}
    &  \Gamma_0=(3k+2)R+(3k^2-3k-2)S-(k+2)T-(3k+2)(k-1)P, \\
      &  \Gamma_1=(3k+2)(k-2)(R-S-T+P).
     \end{split} 
\end{equation}
For the variance, we have
\begin{equation}\label{sec03-eq4}
\begin{split}
var\Big[\Delta p_C\Big|p_C(0)=p\Big]&=\mathbb{E}\Big[(\Delta p_C)^2\Big|p_C(0)=p\Big]-\mathbb{E}\Big[\Delta p_C\Big|p_C(0)=p\Big]^2\\
&=\frac{1}{N^2}\Bigg(\mathbb{P}\Big(\Delta p_C=-\frac{1}{N}\Big)+\mathbb{P}\Big(\Delta p_C=\frac{1}{N}\Big)\Bigg)\Delta t+\mathcal{O}(\delta^2)\Delta t\\
&\simeq \frac{(k-2)}{N^2(k-1)}p(1-p)\Delta t \Bigg(\equiv\nu(p)\Delta t\Bigg).
\end{split}
\end{equation}
Conditions (\ref{sec03-eq2}) and (\ref{sec03-eq4}) ensure the diffusion approximation with drift function $\mu(x)$ and diffusion function $\nu(x)$.

Suppose that a proportion $p$ of cooperators appears in a population of defectors, $p\in(0,1)$. As a result, there are two possibilities for the evolutionary dynamics. The first scenario is that this proportion produces a lineage, which will eventually invade the entire population (extinction of defectors $x=1$). The second scenario is that these proportion might die before reproducing or generate a lineage that disappears after sometime (extinction of cooperators $x=0$). Then, $x=0$ and $x=1$ are absorbing states of the diffusion process.

Let $\phi^{\delta}_C(p,t)$ be the probability that absorption has occurred at $x=1$ at or before $t$. The backward Kolomogov equation (Kimura \cite{K1962}, Crow and Kimura \cite{CK1970}, Ewens \cite{E2004}) can be written as
\begin{equation}\label{fixation}
\frac{\partial \phi^{\delta}_C(p,t)}{\partial t}=\mu(p)\frac{\partial
\phi^{\delta}_C(p,t)}{\partial p}+\frac{\nu(p)}{2}\frac{\partial^2
\phi^{\delta}_C(p,t)}{\partial p^2}
\end{equation}
with boundary conditions $\phi^{\delta}_C(0,t)=0$ and $\phi^{\delta}_C(1,t)=1$.

By letting $t\rightarrow\infty$,  the limit
\begin{eqnarray}
\phi^{\delta}_C(p) = \lim_{t\to\infty}\phi^{\delta}_C(p,t)
\end{eqnarray}
represents the fixation probability of cooperators given an initial frequency $p$. As $t\to\infty$, the left-hand side in (\ref{fixation}) tends to $0$, since $\phi^{\delta}_C(p,t)$ tends to be constant. Therefore, Eq. (\ref{fixation}) becomes
 \begin{equation}\label{sec03-eq5}
\mu(x)\frac{d\phi^{\delta}_C}{dx}(x)+\frac{\nu(x)}{2}\frac{d^2\phi^{\delta}_C}{dx^2}(x)=0,
\end{equation}
with the boundary conditions, $\phi^{\delta}_C(0)=0$ and $\phi^{\delta}_C(1)=1$.
The solution of Eq (\ref{sec03-eq5}) is
\begin{equation}\label{sec03-eq6}
\phi^{\delta}_C(p)=\int_{0}^p \exp\Big\{-\int_{0}^x\frac{2\mu(y)}{\nu(y)}dy\Big\}dx\Big/\int_{0}^1 \exp\Big\{-\int_{0}^x\frac{2\mu(y)}{\nu(y)}dy\Big\}dx.
\end{equation}
Using Eqs (\ref{sec03-eq2}) and (\ref{sec03-eq4}), we have
\begin{equation}\label{sec03-eq7}
\begin{split}
\exp\Big\{-\int_{0}^x\frac{2\mu(y)}{v(y)}dy\Big\}=&\exp\Big\{-\int_{0}^x\frac{\delta N}{2k^2}(\Gamma_1y+\Gamma_0)dy\Big\}\\
=&\exp\Big\{\frac{-\delta N}{4k^2}(\Gamma_1 x^2+2\Gamma_0x+k^2\Delta\alpha x)\Big\}\\
=&1-\delta\cdot\frac{N}{4k^2}\Big[\Gamma_1 x^2+2\Gamma_0x+2k^2\Delta\alpha x\Big]+\mathcal{O}(\delta^2).
\end{split}
\end{equation}
Integrating Eq (\ref{sec03-eq7}), we have the key approximation 
\begin{equation}\label{sec03-eq8}
\begin{split}
\phi^{\delta}_C(p)=&\frac{\int_{0}^p \Big(1-\delta\cdot\frac{N}{4k^2}(\Gamma_1x^2+2\Gamma_0x+2k^2\Delta\alpha x)\Big)dx+\mathcal{O}(\delta^2)}{\int_{0}^1 \Big(1-\delta\cdot\frac{N}{4k^2}(\Gamma_1x^2+2\Gamma_0x+2k^2\Delta\alpha x)\Big)dx+\mathcal{O}(\delta^2)}\\
=&\frac{p-\delta\cdot\frac{N}{4k^2}\Big[\frac{\Gamma_1p^3}{3}+\Gamma_0p^2+k^2\Delta\alpha p^2)\Big]+\mathcal{O}(\delta^2)}{1-\delta\cdot\frac{N}{4k^2}\Big[\frac{\Gamma_1}{3}+\Gamma_0+k^2\Delta\alpha\Big]+\mathcal{O}(\delta^2)}\\
=& \Bigg[p-\delta\cdot\frac{N}{4k^2}\Big(\frac{\Gamma_1p^3}{3}+\Gamma_0p^2+k^2\Delta\alpha p^2\Big)\Bigg]\times \Bigg[1+\delta\cdot\frac{N}{4k^2}\Big(\frac{\Gamma_1}{3}+\Gamma_0+k^2\Delta\alpha\Big)\Bigg]+\mathcal{O}(\delta^2)\\
=&p+\delta\cdot\frac{Np(1-p)}{12k^2}\Big[\Gamma_1+3\Gamma_0+3k^2\Delta\alpha+\Gamma_1p\Big]+\mathcal{O}(\delta^2).
\end{split}
\end{equation}

Similarly, let $\phi^{\delta}_D(p)$ be the probability that a proportion $p$ of defectors takes over a population of cooperators. Since there is ultimate fixation of cooperation or defection with probability $1$, we have 
\begin{equation}\label{sec03-eq9}
\phi^{\delta}_D(p)=1-\phi^{\delta}_C(1-p)=p-\delta\cdot\frac{Np(1-p)}{12k^2}\Big[2\Gamma_1+3\Gamma_0+3k^2\Delta\alpha-\Gamma_1p\Big]+\mathcal{O}(\delta^2).
\end{equation}

Note that the above calculation is valid only for $k\geq 3$ since both the expectation and the variance, given by Eqs (\ref{sec03-eq2}) and (\ref{sec03-eq4}), are zero. As a result of  Eqs (\ref{sec03-eq8}) and (\ref{sec03-eq9}), we have 
\begin{equation}\label{ratio}
\frac{\phi^{\delta}_C(p)}{\phi^{\delta}_D(p)}=1+\delta\cdot\frac{N(1-p)}{4k^2}\Big[\Gamma_1+2\Gamma_0+2k^2\Delta\alpha \Big]+\mathcal{O}(\delta^2).
\end{equation}


\section{Appendix B: Circular model ($k=2$)}
Suppose that we have $N$ sites  over a circle numbered $1,2\ldots,N$. Each site is occupied by an individual. Individual who is located at site $l$ can interact with its neighbors located at sites $l-1$ and $l+1$, through game matrix (\ref{M1}). The same graph is used for the replacement graph. Dissatisfied individuals imitate their direct neighbors.

At each time step, each individual interacts with its direct neighbors. Then, an individual $I$ is chosen at random. It will update its strategy with probability (\ref{prob-imitate}). In this case, it will imitate the strategy of a direct neighbor $J$, with probability proportional to its fitness $f_J=1+\delta \Pi_J$. Otherwise, individual $I$ will keep its current strategy.

The population is initially consisted entirely of defectors. A new cooperator is introduced on a particular site. We have two scenarios. This cooperator will generate a lineage of cooperators without overlapping one beside the other, which will take over the population. In this case, the population ends with only cooperators (extinction of defectors). The second scenario is that this individual might die before reproducing or generate a lineage that disappears (extinction of cooperators). Let $\rho_C(\delta)$ the probability of the first scenario. Likewise, $\rho_D(\delta)$ is the probability that a single defector placed in a population of cooperators will generate a lineage, which will take over the population. Using a recursive argument (Karlin and Taylor \cite{KT1975}), we have
\begin{equation}\label{B-eq1}
\begin{split}
&\rho_C(\delta)=\frac{1}{1+\sum_{i=1}^{N-1}\prod_{j=1}^{i}\frac{T_j^-}{T_j^+}},\\
& \frac{\rho_C(\delta)}{\rho_D(\delta)}=\prod_{i=1}^{N-1}\frac{T_i^+}{T_i^-}.
\end{split}
\end{equation}
where $T_i^+$ (resp. $T_i^-$) is the transition probability of "the number of cooperators increases from $i$ to $i+1$ in one time step" (resp."the number of cooperators decreases from $i$ to $i-1$ in one times step"). 

\subsection{Payoffs}
Without loss of generality, suppose that sites $l+1,\ldots,l+i$ are occupied by cooperators, while the other sites are occupied by defectors. Changes in the composition of the population take place at the boundary between the two clusters: cooperators' cluster formed by sites $l+1,\ldots,l+i$ and defectors' cluster formed by the other sites. Changes in one time step may happen at sites $l,l+1,l+i,l+i+1$. 

To find the transition probabilities $T_i^+$ and $T^-_i$,  the different payoffs of individuals around the boundary should be known. The payoff of an individual depends on the number of its neighbors of each type. We have the following types of payoffs
\begin{align}
&\Pi_{C,(1,1)}=\frac{R+S}{2},\nonumber\\
&\Pi_{D,(1,1)}=\frac{T+P}{2},\nonumber\\
&\Pi_{C,(2,0)}=R,\nonumber\\
&\Pi_{C,(0,2)}=S,\\
&\Pi_{D,(2,0)}=T,\nonumber\\
&\Pi_{D,(0,2)}=P,\nonumber
\end{align} 
where $\Pi_{X,(l,j)}$ refer to the payoff of an $X$-player, who has $l$ cooperators and $j$ defectors as neighbors, for $X\in\{C,D\}$ and $l+j=2$ is the graph degree.

\subsection{Ratio of transition probabilities}
The transition $i\rightarrow i+1$ takes place only if a defector, who is located at the boundary, becomes a cooperator. This occurs with probability
\begin{align}\label{B-eq2}
T_i^+&=\frac{2}{N}\times
\mathbb{E}\Big[\frac{1}{1+e^{\delta(\Pi_{D,(1,1)}-\alpha_D)}}\Big]\times\frac{f_{C,(1,1)}}{f_{C,(1,1)}+f_{D,(0,2)}}\nonumber\\
&=\frac{1}{2N}+\frac{\delta}{4N}\Big[ \Pi_{C,(1,1)}+\tilde{\alpha}_D-\Pi_{D,(0,2)}-\Pi_{D,(1,1)}\Big]+\mathcal{O}(\delta^2).
\end{align}
The transition $i\rightarrow i-1$ takes place only if a defector, who is located at the boundary, becomes a cooperator. This occurs with probability
\begin{align}\label{B-eq3}
T_i^-&=\frac{2}{N}\times
\mathbb{E}\Big[\frac{1}{1+e^{\delta(\Pi_{C,(1,1)}-\alpha_C)}}\Big]
\times\frac{f_{D,(1,1)}}{f_{C,(2,0)}+f_{D,(1,1)}}\nonumber\\
&=\frac{1}{2N}+\frac{\delta}{4N}\Big[ \Pi_{D,(1,1)}+\tilde{\alpha}_C-\Pi_{C,(2,0)}-\Pi_{C,(1,1)}\Big]+\mathcal{O}(\delta^2).
\end{align}
Dividing Eq (\ref{B-eq2}) by Eq (\ref{B-eq3}), we obtain
\begin{equation}\label{B-eq4}
\frac{T_i^+}{T_i^-}=1+\frac{\delta}{2}\Big[ 2(\Pi_{C,(1,1)}-\Pi_{D,(1,1)})+\Pi_{C,(2,0)}-\Pi_{D,(0,2)}+\Delta\alpha\Big]+\mathcal{O}(\delta^2).
\end{equation}
Note that Eq (\ref{B-eq4}) is valid for $i=3,\ldots,N-3.$

For $i=2$, only two cooperators are present in the population. Their payoffs are of type $\Pi_{C,(1,1)}$. As a result, we have
\begin{align}
T_2^-&=\frac{2}{N}\times 
\mathbb{E}\Big[\frac{1}{1+e^{\delta(\Pi_{C,(1,1)}-\alpha_C)}}\Big]
\times\frac{f_{D,(1,1)}}{f_{C,(1,1)}+f_{D,(1,1)}}\nonumber\\
&=\frac{1}{2N}+\frac{\delta}{4N}\Big[ \Pi_{D,(1,1)}-2\Pi_{C,(1,1)}+\tilde{\alpha}_C\Big]+\mathcal{O}(\delta^2).
\end{align} 
The transition probability $T_2^+$ is the same in Eq (\ref{B-eq2}). Then, the ratio becomes
\begin{equation}\label{B-eq5}
\frac{T_2^+}{T_2^-}=1+\frac{\delta}{2}\Big[ 2(\Pi_{C,(1,1)}-\Pi_{D,(1,1)})+\Pi_{C,(1,1)}-\Pi_{D,(0,2)}+\Delta\alpha\Big]+\mathcal{O}(\delta^2).
\end{equation}
Likewise, for $i=N-2$, we have
 \begin{equation}\label{B-eq6}
\frac{T_{N-2}^+}{T_{N-2}^-}=1+\frac{\delta}{2}\Big[ 2(\Pi_{C,(1,1)}-\Pi_{D,(1,1)})+\Pi_{C,(2,0)}-\Pi_{D,(1,1)}+\Delta\alpha\Big]+\mathcal{O}(\delta^2).
\end{equation}
 
Finally, for $i=1$, only one cooperator is in the competition with $N-1$ defectors. If it decides to update its strategy, it will switch its strategy with probability $1$ since its direct neighbors are defecting. Therefore, we obtain
\begin{align}
T_1^-&=\frac{1}{N}\times 
\mathbb{E}\Big[\frac{1}{1+e^{\delta(\Pi_{C,(0,2)}-\alpha_C)}}\Big]\\
&=\frac{1}{2N}+\frac{\delta}{4N}\Big[ \tilde{\alpha}_C-\Pi_{C,(0,2)}\Big]+\mathcal{O}(\delta^2),
\end{align} 
whereas 
\begin{align}
T_1^+&=\frac{2}{N}\times 
\mathbb{E}\Big[\frac{1}{1+e^{\delta(\Pi_{D,(1,1)}-\alpha_D)}}\Big]
\times\frac{f_{C,(0,2)}}{f_{C,(0,2)}+f_{D,(0,2)}}\nonumber\\
&=\frac{1}{2N}+\frac{\delta}{4N}\Big[ \Pi_{C,(0,2)}-\Pi_{D,(0,2)}+\tilde{\alpha}_D-\Pi_{D,(1,1)}\Big]+\mathcal{O}(\delta^2).
\end{align} 
 Accordingly, the ratio becomes
\begin{equation}\label{B-eq7}
\frac{T_1^+}{T_1^-}=1+\frac{\delta}{2}\Big[ 2\Pi_{C,(0,2)}-\Pi_{D,(0,2)}-\Pi_{D,(1,1)}+\Delta\alpha\Big]+\mathcal{O}(\delta^2).
\end{equation}
Likewise, for $i=N-1$, we have
 \begin{equation}\label{B-eq8}
\frac{T_{N-1}^+}{T_{N-1}^-}=1+\frac{\delta}{2}\Big[ \Pi_{C,(2,0)}+\Pi_{C,(1,1)}-2\Pi_{D,(2,0)}+\Delta\alpha\Big]+\mathcal{O}(\delta^2).
\end{equation}

\subsection{Approximation of the fixation probabilities}
We expand  $\prod_{j=1}^{i}\frac{T_j^-}{T_j^+}$ up to the first-order in $\delta$,
\begin{equation}
\prod_{j=1}^{i}\frac{T_j^-}{T_j^+}=\prod_{j=1}^{i}\Big[1+\delta\cdot\frac{d}{d\delta}\Big(\frac{T_j^-}{T_j^+}\Big)\Big|_{\delta=0}\Big]=1+\delta\cdot\sum_{j=1}^{i}\frac{d}{d\delta}\Big(\frac{T_j^-}{T_j^+}\Big)\Big|_{\delta=0}+\mathcal{O}(\delta^2).
\end{equation}
Accordingly, we have
\begin{equation}\label{B-eq9}
\begin{split}
&\rho_C(\delta)=\frac{1}{N+\delta\cdot\sum_{i=1}^{N-1}\sum_{j=1}^{i}\frac{d}{d\delta}\Big(\frac{T_j^-}{T_j^+}\Big)\Big|_{\delta=0}+\mathcal{O}(\delta^2)}\\
&=\frac{1}{N}+\delta\cdot\frac{1}{N}\sum_{i=1}^{N-1}\sum_{j=1}^{i}\frac{d}{d\delta}\Big(\frac{T_j^+}{T_j^-}\Big)\Big|_{\delta=0}+\mathcal{O}(\delta^2)\\
&=\frac{1}{N}+\delta\cdot\frac{1}{N}\sum_{j=1}^{N-1}(N-j)\frac{d}{d\delta}\Big(\frac{T_j^+}{T_j^-}\Big)\Big|_{\delta=0}+\mathcal{O}(\delta^2),\\
&\frac{\rho_C}{\rho_D}(\delta)=1+\delta\cdot\sum_{j=1}^{N-1}\frac{d}{d\delta}\Big(\frac{T_j^+}{T_j^-}\Big)\Big|_{\delta=0}+\mathcal{O}(\delta^2).
\end{split}
\end{equation}
Substituting  Eqs (\ref{B-eq4},\ref{B-eq5},\ref{B-eq6},\ref{B-eq7},\ref{B-eq8}) in Eq (\ref{B-eq9}) yield to
\begin{equation}\label{B-eq10}
\begin{split}
\rho_C(\delta)=&\frac{1}{N}+\delta\cdot\frac{1}{4N^2}\Big[(2N^2-7N+7)R+(N^2+2N-3)S-(N^2-2N+5)T\\
&-(2N^2-3N-5)P+N(N-1)\Delta\alpha\Big]+\mathcal{O}(\delta^2),
\end{split}
\end{equation}
and
\begin{equation}\label{B-eq11}
\frac{\rho_C}{\rho_D}(\delta)=1+\frac{\delta}{2}\Big[(2N-5)(R-P)+N(S-T)+N\Delta\alpha\Big]+\mathcal{O}(\delta^2).
\end{equation}


\section{Appendix C: Well-mixed population}
Consider a well-mixed population of size $N$, where each individual can interact with any other individual with the same  probability through game matrix (\ref{M1}). At any time step, all individuals interact by pairs to accumulate payoffs. Then, an individual $I$ is chosen at random to update its strategy. It will update its strategy with probability (\ref{prob-imitate}). In this case, it imitates individual $J$, one of its neighbors, probability proportional to its fitness $f_J=1+\delta \Pi_J$. Otherwise, the current strategy of individual $I$ will be maintained. Similarly to appendix $B$, we have
\begin{equation}\label{C-eq1}
\begin{split}
&\rho_C(\delta)=\frac{1}{1+\sum_{i=1}^{N-1}\prod_{j=1}^{i}\frac{T_j^-}{T_j^+}},\\
& \frac{\rho_C(\delta)}{\rho_D(\delta)}=\prod_{i=1}^{N-1}\frac{T_i^+}{T_i^-}.
\end{split}
\end{equation}
where $T_i^+$ (resp. $T_i^-$) is the transition probability $i\rightarrow i+1$ (resp. $i\rightarrow i-1$).

Suppose that the population is composed of $i$ cooperators and $N-i$ defectors. Then, the payoffs of a cooperator and a defector are given, respectively, by 
\begin{align}\label{C-eqq1}
\Pi_{C,i}&=\frac{(i-1)R+(N-i)S}{N-1}\nonumber\\
\Pi_{D,i}&=\frac{iT+(N-i-1)P}{N-1}.
\end{align}
$T_i^+$ is the probability that a defector, chosen to update its strategy, becomes a cooperator. This occurs with probability
\begin{align}\label{C-eq2}
T_i^+&=\frac{N-i}{N}\times
\mathbb{E}\Big[\frac{1}{1+e^{\delta(\Pi_{D,i}-\alpha_D)}}\Big]
\times\frac{if_{C,i}}{if_{C,i}+(N-i-1)f_{D,i}}\nonumber\\
&=\frac{(N-i)i}{2N(N-1)}+\delta\cdot\frac{(N-i)i}{2N(N-1)}\times\Big[ \frac{\tilde{\alpha}_D-\Pi_{D,i}}{2}+\frac{N-i-1}{N-1}(\Pi_{C,i}-\Pi_{D,i})\Big]+\mathcal{O}(\delta^2).
\end{align}
$T_i^-$ is the probability that a cooperator, chosen to update its strategy, becomes a defector. This occurs with probability
\begin{align}\label{C-eq3}
T_i^-&=\frac{i}{N}\times 
\mathbb{E}\Big[\frac{1}{1+e^{\delta(\Pi_{C,i}-\alpha_C)}}\Big]
\times\frac{(N-i)f_{D,i}}{(i-1)f_{C,i}+(N-i)f_{D,i}}\nonumber\\
&=\frac{(N-i)i}{2N(N-1)}+\delta\cdot\frac{(N-i)i}{2N(N-1)}\Big[\frac{\tilde{\alpha}_C-\Pi_{C,i}}{2}+ \frac{i-1}{N-1}(\Pi_{D,i}-\Pi_{C,i})\Big]+\mathcal{O}(\delta^2).
\end{align}
Therefore, the ratio of transition probabilities is
\begin{equation}\label{C-eq4}
\frac{T_i^+}{T_i^-}=1+\frac{\delta}{2}\Big[\frac{3N-5}{N-1}(\Pi_{C,i}-\Pi_{D,i})+\Delta\alpha\Big]+\mathcal{O}(\delta^2).
\end{equation} 
 Inserting Eq (\ref{C-eq4}) in Eq (\ref{C-eq1}) after expanding them up to the first-order in $\delta$, we have
\begin{align}\label{C-eq5}
\rho_C(\delta)&=\frac{1}{N}+\delta\cdot\frac{1}{N}\sum_{i=1}^{N-1}(N-i)\frac{d}{d\delta}\Big(\frac{T_i^+}{T_i^-}\Big)\Big|_{\delta=0}+\mathcal{O}(\delta^2)\nonumber\\
&=\frac{1}{N}+\delta\cdot\frac{N-1}{4}\Bigg[\frac{3N-5}{3(N-1)^2}\Big((N-2)R+(2N-1)S-(N+1)T-(2N-4)P\Big)+\Delta\alpha\Bigg]
\end{align}
and
\begin{align}\label{C-eq6}
\frac{\rho_C}{\rho_D}(\delta)&=1+\delta\cdot\sum_{i=1}^{N-1}\frac{d}{d\delta}\Big(\frac{T_i^+}{T_i^-}\Big)\Big|_{\delta=0}+\mathcal{O}(\delta^2)\nonumber\\
&=1+\delta\cdot\frac{N-1}{2}\Bigg[\frac{3N-5}{2(N-1)^2}\Big((N-2)R+NS-NT-(N-2)P\Big)+\Delta\alpha\Bigg]+\mathcal{O}(\delta).
\end{align}
Note that Eqs (\ref{C-eq5}) and (\ref{C-eq6}) is valid for any finite population size $N\geq 2$.

\newpage

\end{document}